\documentclass{emulateapj}
\usepackage{apjfonts}
\usepackage{color}

\newcommand\beq{\begin{equation}}
\newcommand\eeq{\end{equation}}

\shorttitle{SGRBs from BH-BH Mergers}
\shortauthors{Perna, Lazzati \& Giacomazzo}

\begin{document}

\title{Short Gamma-Ray Bursts from the Merger of Two Black Holes}

\author{Rosalba Perna\altaffilmark{1}, Davide Lazzati\altaffilmark{2}, Bruno
  Giacomazzo\altaffilmark{3,4}}

\affil{$^1$Department of Physics and Astronomy, Stony Brook University, Stony Brook, NY, 11794, USA}
\affil{$^2$ Department of Physics, Oregon State University, 301
Weniger Hall, Corvallis, OR 97331, USA}
\affil{$^3$ Physics Department, University of Trento, via Sommarive 14, I-38123 Trento, Italy}
\affil{$^4$ INFN-TIFPA, Trento Institute for Fundamental Physics and Applications, via Sommarive 14, I-38123 Trento, Italy}

\begin{abstract}
Short Gamma-Ray Bursts (GRBs) are explosions of cosmic origins believed to be
associated with the merger of two compact objects, either two neutron
stars, or a neutron star and a black hole. The presence of at least
one neutron star has long been thought to be an essential element of
the model: its tidal disruption provides the needed baryonic material
whose rapid accretion onto the post-merger black hole powers the
burst.  The recent tentative detection by the {\em Fermi} satellite of a short
GRB in association with the gravitational wave signal GW150914
produced by the merger of two black holes has {  challenged} this standard
paradigm. Here we show that the evolution of two high-mass,
low-metallicity stars with main sequence rotational speeds a few tens of
percent of the critical speed eventually undergoing a weak supernova
explosion {\em can} produce a short gamma-ray burst.  The outer layers
of the envelope of the last exploding star remain bound and
circularize at large radii. With time, the disk cools and becomes
neutral, suppressing the magneto-rotational instability, and hence the
viscosity. The disk remains 'long-lived dead' until tidal torques and
shocks during the pre-merger phase heat it up and re-ignite accretion,
rapidly consuming the disk and powering the short gamma-ray burst.
\end{abstract}


\section{Introduction}

The recent detection of gravitational waves (GW) from the merger of a
massive stellar binary black hole system (BH, \citealt{abbott2016})
has opened a new window for the observation of the universe. The
gravitational wave signal was determined to be produced by the final
inspiral and ringdown of a binary system of two BHs of masses $36^{+5}_{-4}$ and
 $29^{+4}_{-4}M_\odot$, at a distance of 0.4 Gpc from Earth. The Gamma-Ray Burst
monitor (GBM) on-board Fermi was serendipitously pointing, at that
time, to a region of the sky that intersected with more than 70\% of
the error region from which the GW signal was detected. Analysis of
the data revealed the presence of a mildly significant source of hard
X-rays/soft gamma-rays 0.4 seconds after the GW event, lasting
approximately 1 second (\citealt{connaughton2016}). The  chance
significance of the transient was {  estimated to be} 0.0022. 

If true, the new source would have properties resembling a weak
short-duration Gamma-Ray Burst (GRB). Its isotropically equivalent
energy in the GBM band would amount to
$L_{\rm{iso}}\simeq2\times10^{49}$~erg~s$^{-1}$, about a factor 10
weaker than a typical short GRB (e.g. Li et al. 2016), but not unprecedented (e.g.,
GRB080925A, \citealt{davanzo2014}). While the intrinsic weakness of
the burst energetics could suggest that the short GRB was seen
off-axis and its true energy is much larger, the hardness of the
spectrum requires instead that the burst is seen on-axis and the
measured energy corresponds to the true energy budget of the event. If
not due to a random fluctuation, the Fermi observation points
therefore to the association of weak, mildly (if at all) collimated
short GRBs with the merging of BH-BH binary systems. 

Such association is somewhat surprising since general consensus
requires the presence of at least one neutron star (NS) in the merging
compact binary system that generates a gamma-ray bursts
(\citealt{berger2014}). This requirement is motivated by the need of
creating a post-merger BH-accretion disk system. In a NS-NS or NS-BH
merger scenario, the NS(s) would be tidally stripped of
$\sim0.1\,M_\odot$ of matter during the merger and provide the
material for an accreting circum-BH torus
(\citealt{rezzolla2010,Giacomazzo2013}). In a clean, double BH merger
case, the source of the accreting material was expected to be absent,
{  which is why the {\em Fermi} data, albeit at low statistical
  significance, have triggered so much attention. However,
  we should point out that, following those observations,
  {\em INTEGRAL} imaged the same field, but did not identify any
  source up to a limit on the $\gamma$-ray isotropic equivalent
  luminosity of $E_\gamma < 2\times 10^{48}$~erg, hence questioning
  the association (Savchenko et al. 2016).}

Whether the association between GW150914 and the {\em Fermi} GBM
candidate is real or not, it has raised the possibility that BH-BH
mergers may also produce SGRBs, unlike commonly thought. In this
letter we show that such a scenario is possible within the standard
theory of high mass evolution and accretion disks.

\section{Formation of a binary BH-BH with a fallback disk}

In the following, we consider a typical evolutionary scenario that may
lead to electromagnetic radiation in conjunction with a BH-BH
merger. As a specific example, we adopt parameter values that would
lead to a type of event such as GW150904, but clearly our
considerations can be generalized to other initial conditions, and
hence the specific model discussed here should simply be considered as
a representative case of the new idea that we are proposing.

The starting point for this representative case is a high-mass binary
system with stars of masses in the 30 to 40~$M_\odot$ range and
metallicity $Z\la 0.1 Z_\odot$. If the two stars were to evolve
without interaction, such as in a detached binary, then, for this
range of initial masses, a BH is generally expected to be formed
either through a partial or a full fallback of the envelope
(e.g. Fryer \& Kalogera 2001; Heger et al. 2003). However, there are
several other key properties of the stars beyond their initial masses,
as well as elements of stellar evolution, which greatly influence this
simple picture.  In particular, metallicity plays a fundamental role
in mass loss, since a higher metallicity results in stronger
radiation-driven winds (e.g. {  Vink et al. 2001; Mokiem et
  al. 2007}), and at high metallicities NSs {  are predicted to}
become an increasingly more common remnant {  (e.g. Heger et
  al. 2003)}. Additionally, the initial rotational velocity of the
star also affects evolution, since it influences elemental mixing
within the envelope, as well as angular momentum transport (e.g. De
Mink et al. 2009). The latter, as discussed more in the following, is
also highly dependent on the uncertain role played by magnetic torques
during the evolution of the star.

When stars evolve in a binary system, in addition to the
elements discussed above, also binary interactions play an important
role, and many studies have been devoted to model these interactions
with various degrees of sophistication (for a review of the channels
forming a double BH system, see e.g. Kalogera et al. 2007). If the
stars are rapidly rotating and remain chemically homogeneous and
compact throughout their lifetime without becoming giants, then they
may evolve {  without exchanging mass} (e.g. de Mink et al. 2009).  However, in
close binary systems in which stars evolve to the giant phase, the
exchange between the two stars can be dramatic, including
non-conservative mass transfer (e.g. Dominik et al. 2012).

However, despite these interactions, the evolution of the mass gainer
star after the accretion episode is found to be almost identical to
that of a star evolving in isolation with the same mass ({  Braun et
al. 1995; Dray et al. 2007}).  In particular, these authors find that the angular
momentum is very similar in the isolated and in the binary evolution
cases. These similarities arise because, while the post-accretion,
rejuvenated star is a bit more evolved than the isolated one, however
this difference is small compared to the longer secular evolution, and
it mainly result in a slightly reduced loss of spin down due to mass
loss during core hydrogen burning.

Given the above, and {  the limited studies} of the pre-supernova
(SN) interior structure of rotating stars in interacting binaries
{  (e.g. Yoon et al. (2010) studied stars with initial mass of the
  binary component in the 12-25 $M_\odot$ range)}, here we consider
some examples of the pre-SN structure of rotating, isolated stars.  We
emphasize that these models should be simply considered as
illustrative of the conditions required in the interior of the star
prior to its collapse for our scenario to work.

We have calculated the pre-SN star models using the evolutionary code
MESA\footnote{  We used version 4770. The inlist can be 
downloaded from http://www.astro.sunysb.edu/rosalba/inlist.} 
inclusive of rotation (Paxton et al. 2013).  {  The pre-SN
  star is identified at the time of core collapse, which in turn is
  defined as the moment when any part of the Fe core is falling with a
  velocity $\ga 1000$~km~s$^{-1}$. } An important point to note is
that these simulations include angular momentum transport via magnetic
torques (Spruit 2002). However, the importance of these torques is
still a subject of debate, and it may very well vary considerably from
star to star.  If magnetic torques are negligible during the
evolution, then the specific angular momentum of the pre-SN star is
considerably higher for the same initial conditions (see e.g. fig.4 of
Perna et al. 2014), and therefore the requirement of low metallicity
in our models would become less stringent.

\begin{figure}
\centering 
\vspace{-0.5cm}
\includegraphics[width=0.495\textwidth]{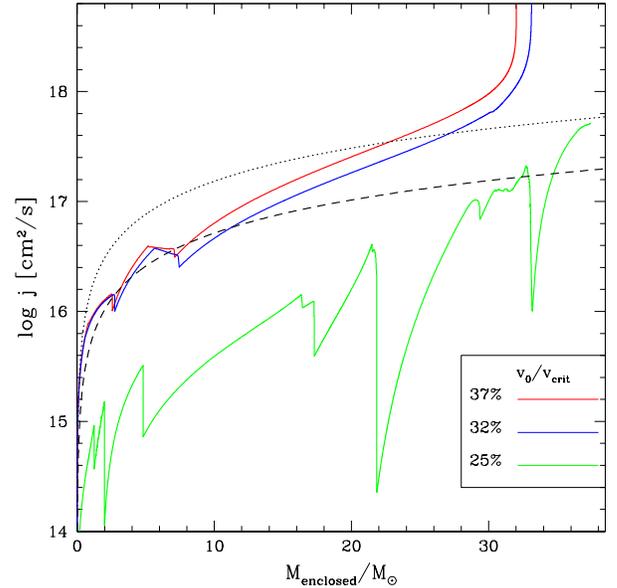}
\vspace{-2.cm}
\caption{The distribution of specific angular momentum in a pre-SN
  star of mass $M=40~M_\odot$ and metallicity $Z=0.01Z_\odot$, for
  three initial values of the surface equatorial velocity of the main
  sequence star, expressed in units of the critical surface equatorial
  velocity.  Also plotted is the
  specific angular momentum of a particle at the last stable orbit
  around a Schwarzschild (dotted line) and a Kerr (dashed line) black
  hole of mass $M=M_{\rm enclosed}$.}
\label{fig:fig1}
\end{figure}

Fig.~1 shows the distribution of the specific angular momentum in the
interior of the star just prior to its explosion, for a main sequence star of 
mass $M=40~M_\odot$, metallicity $Z=0.01~Z_\odot$, and three
values of the initial equatorial rotational speed.  Also
indicated is the specific angular momentum of a particle in a
corotating orbit at radius $R$ around a BH of spin parameter $a=0$
(dotted line) and $a=1$ (dashed line), where {  (e.g. Shapiro \& Teukolsky 1983)} 
\begin{equation} j(R) = \frac{\sqrt{GMR} \left[R^2 -
    2(a/c)\sqrt{GMR/c^2} +(a/c)^2\right]} {R\left[ R^2-3GMR/c^2 +
    2(a/c)\sqrt{GMR/c^2}\right]^{1/2}}\;.
\label{eq:jGR}
\end{equation} 
Here $M$ is the mass of the BH and $J=aM$ its angular momentum. 

The pre-SN models of Fig.~1 are characterized by the fact that the
outer envelope layers are endowed with a specific angular momentum
$j_m >j(R_{\rm lso})$, where $R_{\rm lso}$ is the radius at the last
stable orbit.  This is a structure that we envisage in at least one of
the two stars, preferably the second one to go off as SN, since any
fallback material left over from the first SN may be blown away when
the second star explodes. {  The models chosen here, with 
an initial metallicity of $Z=0.02$, have been
  selected to straddle a range of critical velocities over which the
  star transitions from an evolution with strong compositional
  gradients and a red-giant phase, to a chemically-homogeneous
  evolution which bypasses the core-envelope structure.  The critical
  surface equatorial velocity $v_{\rm crit}$ is defined as $v^2_{\rm
    crit}=( 1\,-\,{L}/{L_{\rm Edd}})\,({G\,M}/{R})$, where $R$ is the
  radius of the star of mass $M$, and $L_{\rm Edd}=4\pi cGM/\kappa$,
  with $\kappa$ the opacity. Mass loss to winds is implemented
  following the prescription of Yoon et al. (2006), which includes a
  metallicity scaling of line-driven winds. 
 Exponentially decaying
  overshoot at the convective boundaries is included with a mixing
  parameter $f=0.005$ (Paxton et al. 2011). We used a standard 25
  isotopes nuclear network (\textsc{approx25.net}).  Transport of
  angular momentum and chemicals due to rotationally induced
  instabilities is implemented in a diffusion approximation
  (e.g. Heger et a. 2000). Convection and the Eddington-Sweet
  circulations are believed to be the dominant mixing processes during
  the main sequence of rotating massive stars (e.g. Paxton et
  al. 2013). In the models we initialized rotation and define t=0
  (zero age main sequence) when the total nuclear luminosity generation is at least 95\%
  of the total stellar luminosity.}

Before continuing, we should note that the idea that SN explosions may
leave behind long-lived disks has a long history in the literature,
dating as far back as Colgate (1971) and Chevalier (1989). However,
the focus has always been on fallback disks around NSs rather than
BHs, since the former display a wide range of observable phenomena
which can be accounted for with such disks, from enhanced emission in
the Anomalous X-ray Pulsars (Chatterjee et al. 2000; Alpar 2001), to
anomalous braking indices in pulsars (Menou et al. 2001), to jets in
pulsars (Blackman \& Perna 2002), to transient pulsars (Cordes \&
Shannon 2008), to the making of planets (Lin et al. 1991).  These
long-lived fallback have been shown to be observable up to ages
$\sim 10^4-10^5$~years at long wavelengths, especially the infrared
(Perna et al. 2000), and a detection of a fallback disk has indeed
been made (Wang et al. 2006).  In the context of BHs, the focus has
traditionally been on the rapidly accreting, short-lived disks that
power the long GRBs, since these are connected to a well established
observational phenomenology. However, the possibility of long-lived
disks also around BHs has been introduced by Perna et al. (2014). While
the motivation there was to explore the possibility of planet
formation around BHs, those calculations provide the context for
forming a short GRB during a BH-BH merger, as it will be argued in the
following.

In order to leave behind a BH of mass $\sim 30 M_\sun$, stars such as
the ones considered in Fig.1 need to undergo a relatively weak
explosion.  
For example, a star with $M=40M_\odot$, $Z=0.01Z_\odot$
and $v=37\%$ of the critical speed, would experience $M\ga 30 M_\odot$
of mass in fallback for explosion energies $E\le 10^{51}$~erg (see
fig.6 in Perna et al 2014).  The outer layers of the envelope of these
stars, endowed with angular momentum $j_m>j(R_{\rm los})$, will fall
back on a dynamical timescale and eventually circularize at the radius
$R_{\rm circ}$ where $j_m=j(R_{\rm circ})$.  The following evolution of
the ring is mediated by viscosity, with a typical timescale
\begin{equation}
t_0\,(R_{\rm circ})\,=\,\frac{R_{\rm circ}^2}{H^2 \alpha \Omega_K}\,
\sim 160\; \alpha^{-1}_{-1}\;m^{-1/2}_{30}\,
R_{10}^{3/2}\left(\frac{R}{H}\right)^{2}\,{\rm s} \;,
\label{eq:tvisc}
\end{equation}
where $R_{10}=R/(10^{10}~{\rm cm})$, $m_{30}=M/(30~M_\odot)$, $H$ is
the disk scale-height, $\Omega_K$ is the Keplerian velocity of the gas
in the disk, and $\alpha$ the viscosity parameter in units of
$\alpha_{-1}\equiv\alpha/0.1$ (Shakura \& Sunyaev 1973).
The disk height is $H\sim R$ during the early, hot super-Eddington slim disk phase,
whereas it is $H\sim 0.1~R$ once the disk,
cooled by photons, becomes optically thick, geometrically thin. 
After a time $t\sim t_0$, the evolution of the disk accretion rate 
can be well approximated with a power law (e.g. Cannizzo et al 1990),
\begin{equation}
\dot M_d (t)  =  \dot M_d (t_0) \left( \frac{t}{t_0} \right)^{-\beta}\,\\
\label{eq:evol1}
\end{equation}
where $\beta=4/3$ and $\beta=19/16$ apply during the early 
slim disk phase and during the later geometrically thin regime, respectively.

As a specific, quantitative example, let us now consider the evolution
of one of the pre-SN models shown in Fig.1, and in particular, the
case with main sequence rotational speed equal to 37\% of the critical
speed.  Assuming a weak explosion so that all the material falls back,
we find that the outermost $\sim 0.5~M_\odot$ of envelope circularizes
at radii between $\sim 10^9-5\times 10^{10}$~cm. {  While we focus
  on this fallback material from the pre-SN star for the following
  discussion, we note that there could be substantial more mass at
  large radii ejected with the wind and which survives the SN
  explosion, especially if weak and asymmetric. This outer envelope
  mass is endowed with higher specific angular momentum than the one
  of the pre-SN star (Paxton et al. 2013), and hence, if not unbound
  by the explosion, would easily circularize around the BH. In the following, to
  remain conservative, we limit our discussion to the evolution of the
  fallback mass alone. If there were more mass around, it would only
  make our model requirements less restrictive.} For the particular
model under consideration, the initial accretion rate is on the order
of $\dot{M(t_0)}\sim M_d/t_0 \sim 6\times 10^{-6}M_\odot$~s$^{-1}$.
The disk remains super-Eddington for about 90~years, and continues to
cool with the accretion rate dropping and the mass gradually
depleting.

Accretion can proceed as long as the temperature in the disk remains
high enough to maintain the gas at least partially ionized. However,
as the temperature drops and the magnetic diffusivity decreases,
accretion becomes choked.  Numerical simulations of the
magneto-rotational instability (MRI, Balbus \& Hawley 1991) have
shown that, at low magnetic Reynolds numbers ($\sim$~a few $\times
10^3$) MHD turbulence and its associated angular momentum transport
are significantly reduced (Hawley et al. 1996; Fleming,
et al 2000). The power law evolution of the SN
fallback disk gets interrupted, and the disk becomes 'dead'\footnote{Note that such a type of disk
was detected around an isolated NS, with an estimated age $\ga 10^6$~years (Wang et al. 2006).}
(e.g. Menou et al. 2001). The precise value of the accretion rate
at which this happens depends on the specifics of the opacity, and hence
the composition of the pre-SN star. 
For a disk of solar composition, the local stability criterion is (Hameury et al. 1998)
\begin{equation}
\dot M_d(R) > \dot M_{\rm crit}(R) \simeq 9.5\times 10^{15}\,
m_{10}^{-0.9}\,R_{10}^{2.68}\, {\rm g}~{\rm s}^{-1}\,
\label{eq:Mcrit}
\end{equation}
and slight variations are expected in the case of helium-rich and
metal-rich disks (Menou et al. 2001).  For the representative model
discussed here, when the disk becomes 'dead' it still retains $\sim
5.5\times 10^{-4}M_\odot$ of its mass. Note that this is sufficient to power
a GRB with observed luminosity $L_{\rm iso}\sim 2\times 10^{49}$~erg~s$^{-1}$
such as the one measured for the possible counterpart of GW150914. 

From this point on, if the
binary is not perturbed by external elements, it will live a long time
as a system of two BHs with one surrounded by an inactive disk, until
the final plunge rekindles the disk, as discussed in the following.

\section{The final seconds: making a SGRB}

\begin{figure}
\includegraphics[width=\columnwidth]{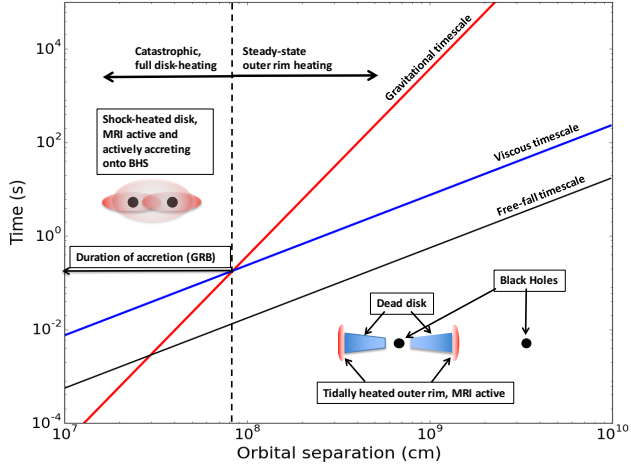}
\caption{{Comparison of the free-free, viscous, and gravitational
    inspiral timescales as a function of the orbital separation for a
    system of two $M=30$~M$_\odot$ black holes. One of the two BHs is
    assumed to be surrounded by a 'dead' fallback disk. The disk is reactived once
the gravitational timescale becomes smaller than the viscous one. From that point
on the two BHs merge on the very short timescale $t_{\rm GW}$, followed by an
electromagnetic emission on the timescale $t_{\rm visc}$.}
\label{fig:timescales}}
\end{figure}

Let us now consider the evolution of a binary black hole system with a
'dead' accretion disk surrounding one of the two black holes
(analogous considerations hold for the case in which both BHs have
accretion disks). If the outer radius of the accretion disk is smaller
than the tidal truncation radius, $R_\mathrm{TT}$, the disk and the
companion BH do not interact significantly\footnote{  Particles
  orbiting outside the tidal radius are more significantly affected by
  the presence of the companion BH, whose tidal effects would cause
  their orbits to be perturbed and intersept each other, in absence of
  any form of viscosity (\citealt{papaloizou1977}).}
(\citealt{paczynski1977,papaloizou1977,ichikawa1994}; {  see also Armitage \& Natarajan 2002 
and Cerioli et al. 2016 for numerical simulations of the 'tidal-squeezing' effect }).  We focus here
on a binary black hole system with two identical black holes and with
orbital separation $r$. We also assume that the disk and the binary
orbits are in the same plane, even though a different geometry should
not affect the conclusions of this argument. The tidal truncation
radius in this case is $R_\mathrm{TT}\sim0.3 R$
(\citealt{paczynski1977}). For any reasonable parameter set, the
viscous timescale at the outer rim of the disk (Eq.~\ref{eq:tvisc}) is
much shorter than the gravitational waves inspiral
timescale\footnote{  We note that the presence of a disk around one of the
  BHs will generally influence the angular momentum of the
  binary, and hence the merger timescale; however, the effect is
  expected to be significant only if the mass of the disk is at least
  comparable with that of the companion BH (Lodato et al. 2009). }
$t_\mathrm{GW}$ (\citealt{huges2009}; see
Figure~\ref{fig:timescales}):
\begin{equation}
t_\mathrm{GW}=\frac{5}{256}\frac{c^5}{G^3}\frac{R^4}{2m^3}
=0.37\frac{R_8^4}{m_{30}^3}~{\rm s}. 
\label{eq:GW}
\end{equation}
In this regime, the bare black hole excites tidal dissipation,
concentrated in the outer rim of the accretion disk
(\citealt{papaloizou1977,ichikawa1994}). The associated heating
ionizes the outer rim of the disk turning on the MRI. Because the
inner part of the disk is still neutral, the material in the outer rim
cannot accrete, and hence piles up at the outer edge of the dead zone. 

As long as $t_\mathrm{GW}>t_0$, the system evolves in a quasi
steady-state fashion, since the disk has time to adjust to the new
BH-BH configuration, maintaining an MRI active outer rim pushing
against an inactive and non-accreting inner disk. As the binary
shrinks, it reaches a point at which $t_\mathrm{GW}\simeq{t_0}$. From
that moment on, the disk does not have time to adjust to the inspiral
of the binary system and the tidal heating reaches the inner part of
the disk, likely becoming an impulsive, shock-driven event rather than
a quasi-stationary process, analogously to what seen in numerical
simulations of extended disks surrounding a central binary BH
(\citealt{farris2015}). 

The critical radius $r_\mathrm{crit}$ at which the two time-scales are
equal is readily derived from Eqs.~\ref{eq:tvisc} and~\ref{eq:GW}:
\begin{equation}
r_{\mathrm{crit}}=3.45\times10^7\,\left(\frac{R}{H}\right)^{4/5}\,
\frac{m_{30}}{\alpha_{-1}^{2/5}}~{\rm cm}.
\end{equation}
The accretion phase is very rapid, since the disk is
very compact due to the accumulation of material at the outer rim that
took place during the inspiral. If accretion produces the launching of
a relativistic jet -- as seen in SGRBs (\citealt{berger2014}) and in
tidal disruption events (\citealt{burrows2011}) -- and the
relativistic jet radiates in gamma-rays, we can derive the burst
duration from the viscous timescale at the critical radius, obtaining:
\begin{equation}
t_\mathrm{GRB}=0.005\,\left(\frac{R}{H}\right)^{16/5}\,
\frac{m_{30}}{\alpha_{-1}^{8/5}}~{\rm s}\,.
\label{eq:tgrb}
\end{equation}

For a relatively thin disk with, e.g.  $(R/H)\sim 3$ at the tidal truncation
radius, Eq.~\ref{eq:tgrb} yields $t_\mathrm{GRB}=0.2$~s, in good
agreement with the Fermi transient associated to GW150914. The burst
luminosity depends on the mass accretion rate, which in turns depends
on the mass of the disk. A disk with a modest mass of
$\sim 10^{-4}-10^{-3}$~M$_\odot$, such as the one discussed in Sect.~2, would
be consistent with the observed luminosity for standard $\sim10\%$
efficiency values for the conversion of accretion power to
relativistic outflow and of the outflow power into radiation.

{  Before concluding, we note that an important condition of our
  model is that the inner disk, say $R_{\rm in}\equiv R(t_{\rm
    GW}=t_0)\lesssim 10^8$~cm, remains cold as long as $t_{\rm
    GW}>t_0$. A potential disturbance may come from the heated outer
  rim, which may produce ionizing photons able to heat and ionize the
  inner regions.  In the following, we estimate the magnitude of such
  a contribution.  Let us consider the binary to be at a separation
  $R$. The outer radius of the disk is then at $R_d=R_{\rm TT}\sim 0.3
  R$. The accretion luminosity is $L_{\rm acc}=\eta GM\dot{M_d}/2R_d$,
  where $\eta$ is an efficiency factor (from mass to radiation), and
  $\dot{M_d}\sim M_d/t_{\rm GW}$. We obtain $L_{\rm acc}\,=\,3\times
  10^{39} \eta\, M_{d,-4}\,m_{30}^3/R^5_{10}$~erg~s$^{-1}$, where
  $M_{d,-4}\equiv M_d/(10^{-4}M_\odot)$. We note that the accretion
  luminosity becomes sub-Eddington at $R_{10}\gtrsim 1$, and is
  limited by the Eddington value $L_{\rm E}=3.7\times
  10^{39}m_{30}$~erg~s$^{-1}$ for $R_{10}\lesssim 1$.  The number of
  ionizing photons is $N_{phot}\propto L\,t_{\rm GW}$, and it drops
  rapidly as $R^{4}$ with orbital separation, for radii
  $R_{10}\lesssim 1$. Hence, for a conservative estimate, we analyze
  the situation at $R_{10}\sim 1$.  Let's assume a typical efficiency
  $\eta\equiv 0.1\eta_{-1}$. The precise spectral shape of the rim is
  not well known, hence we parametrize as $\epsilon_{\rm ph}\equiv 0.1
  \epsilon_{\rm ph,-1}$ the fraction of ionizing photons (UV).  The
  emission geometry is also quite uncertain, hence for simplicity we
  consider it isotropic, and we put ourselves in the most conservative
  case by assuming that the photons are impinging from the rim to the
  inner disk perpendicularly. Then the only reduction is the geometric
  factor $(R_{in}/R_{\rm d})^2=9 (R_{in}/R)^2=9\times 10^{-4}
  (R_{in,8}/R_{10})^2$, accounting for the fraction of photons from
  the rim impacting the inner disk.  Including these factors, the
  number of photoionizing photons is $N_{\rm ph}=L_{\rm
    acc}/(h\nu_{13.6eV})\,t_{\rm GW} \sim 6\times
  10^{52}\eta_{-1}\epsilon_{\rm ph,-1}\,m_{30}\, M^{\rm
    rim}_{-4}\,R_{\rm in,8}^2/R^3_{10}$, having indicated with $M^{\rm
    rim}_{-4}$ the mass in the rim in units of $10^{-4}M_\odot$. 
 This needs to be compared with the number of Hydrogen
  atoms in the inner disk, $N_{\rm H}=M^{\rm inner- disk}/m_p\simeq
  10^{53}\,M^{\rm inner-disk}_{-4}$. We can hence define a parameter
  $\zeta\, =\, 0.6\, \eta_{-1}\,\epsilon_{\rm ph,-1}\,m_{30}\, M^{\rm
    rim}_{-4}\,\,R_{\rm in,8}^2\,R^{-3}_{10}\,(M^{\rm
    inner-disk}_{-4})^{-1}$, with the understanding that it represents
  the fraction of the inner disk which could be ionized by the rim
  prior to the final merger. In order to have the major output of the
  accretion energy {\em following} the merger, this quantity has to be
  small. Otherwise, it would result in a longer, lower-luminosity
  event {\em preceeding} the merger. }

\section{Summary}

The discovery of gravitational waves (\citealt{abbott2016}) has opened a
new window on the Universe, and the combined detection of GWs and EM
radiation would enormously increase their diagnostic power. The
possible detection by {\em Fermi} of a short GRB-like counterpart to
GW150914 has been puzzling in light of the fact that no EM emission was
expected from a double BH merger.

{  Whether this particular association is real or not, it has
  however opened an arena for new ideas to be tested with future
  dedicated simulations and observations of other GW events. In this spirit,}  
here we have presented a new  scenario which, starting from a binary
system of two massive, low-metallicity stars, leads to two massive
BHs, (at least) one of which is surrounded by a fallback disk at large
radii. As the disk cools it eventually becomes neutral, the MRI is
suppressed and the disk can then survive for a very long time as a
'dead' disk.  Eventually, when the two BHs start their final dance
towards the inexorable merger, tidal torques and shocks heat up the
gas as the naked BH spirals inward plowing its way within the disk.
Accretion resumes from the outer regions towards the inner ones, and
the mass pile up propagates inwards as the inner parts of the disk get
gradually revived.  Immediately following the merger the disk is fully
revived, and the mass piled up hence accretes very rapidly giving rise
to a short Gamma-Ray Burst.

\acknowledgments 
RP acknowledges support from NASA/{\em Swift} under grant NNX15AR48G.

\end{document}